\documentclass[aps,preprint,showpacs,preprintnumbers,amsmath,amssymb]{revtex4}

\usepackage{dcolumn}
\usepackage{mathrsfs}
\usepackage{bm}
\usepackage{amsmath,amssymb,epsfig,float}

\begin{document}

\title{Second-order phase transition of Kehagias-Sfetsos black hole in  deformed H\v{o}rava-Lifshitz gravity}
\author{Mengjie Wang}
\author{Songbai Chen} \author{ Jiliang {Jing}\footnote{Corresponding author, Electronic address:
jljing@hunnu.edu.cn}}
 \affiliation{ Institute of Physics and
Department of Physics,
Hunan Normal University, Changsha, Hunan 410081, P. R. China \\
and
\\ Key Laboratory of Low Dimensional Quantum Structures and
Quantum Control of Ministry of Education, Hunan Normal University,
Changsha, Hunan 410081, P.R. China}

\vspace*{0.2cm}
\begin{abstract}
\vspace*{0.2cm}

We study the second-order phase transition (SOPT) for the
spherically symmetric Kehagias-Sfetsos (KS) black hole in the
deformed H\v{o}rava-Lifshitz gravity by applying the methods of
equilibrium and non-equilibrium fluctuations. We find that, although
the KS black hole has only one mass parameter as the usual
Schwarzschild ones, the SOPT will take place if the mass of the KS
black hole changes across the critical point
$\frac{\sqrt{5+\sqrt{33}}(\sqrt{33}-1)} {16\sqrt{\omega}} $. The
result show us that there is difference between the
H\v{o}rava-Lifshitz gravity and the Einstein's gravity theory.

\vspace*{1.5cm}


\vspace*{0.5cm}

\end{abstract}

\maketitle

\section{INTRODUCTION}
\vspace*{0.3cm}
\parindent=10pt

It is well known that the non-renormalizable of gravity is a large
challenge for Einstein's gravity theory. Recently, H\v{o}rava
\cite{ho1,ho2,ho3} proposed a new class of quantum gravity which is
non-relativistic and power-counting renormalizable. The key feature
of this theory is that the space and time exhibit Lifshitz scale
invariance $t\rightarrow b^z t$ and $x_i \rightarrow b^z x_i$ with
$z > 1$. It is this anisotropic rescaling that makes H\v{o}rava's
theory power-counting renormalizable. In the IR region such a
H\v{o}rava's theory can reduce to the well-known Einstein gravity.
Therefore, a lot of attention has been focused on this gravity
theory
\cite{KS,cal,TS,muk,Bra,pia,gao,LMP,KK,CY,CCO,MK,CCO1,CJ1,CJ2,RAK1}.

In general, the IR vacuum in Ho\v{r}ava's theory is anti de Sitter
(AdS) spacetimes. In order to obtain a Minkowski vacuum in the IR
sector, one can add the term ``$\mu^4R$" in the action and then take
the $\Lambda_W \to 0$ limit. This does not change the UV properties
of the theory, but it alters the IR properties. Making use of such a
deformed action, Kehagias \textit{et al} \cite{KS} obtain the
asymptotic flat spherically symmetric vacuum black hole solution
which has two event horizons. The heat capacity is positive for the
small KS black hole and it is negative for the large one. It means
that the small KS black holes are the stable in the Ho\v{r}ava's
theory.  The result imply that there exists the distinct differences
between the KS black hole and the Schwarzschild black hole.

The investigation of SOPT of black hole is helpful to explore the
black hole's property\cite{Hooft85}-\cite{Jing02}. To the best of
our knowledge, there is no report about investigation for SOPT of
the KS black hole in the deformed H\v{o}rava-Lifshitz gravity. On
the other hand, there is a paradox about where the SOPT is taken
place for a long time. Some authors
\cite{Davies77,Lau94a,Lau94b,Lau94c} argue that the SOPT is taken
place when $C\rightarrow\infty$ by applying thermodynamical
equilibrium fluctuations. The other \cite{Pavon} calculated the
non-equilibrium fluctuation of mass and entropy and found that these
fluctuations diverge when $r_+\rightarrow r_-$ and they are finite
when $C\rightarrow\infty$. So they put forwards that the SOPT of KS
black hole takes place  where $r_+\rightarrow r_-$ rather than where
$C\rightarrow \infty$. In this paper, we will address these question
carefully by studying the SOPT of the KS black hole \cite{KS} in the
deformed H\v{o}rava-Lifshitz gravity using the methods of
equilibrium and non-equilibrium fluctuations.

The paper is organized as follows. In  Sec. II, we give a brief
description of solution in the deformed H\v{o}rava-Lifshitz black
hole spacetime. In Sec. III, we calculate a SOPT point by using the
method of equilibrium fluctuations. In Sec. IV, we make use of the
non-equilibrium thermodynamic fluctuations and find a SOPT point
that is just the same point as we find in Sec. III. We present our
conclusions and make some discussions in the last section.
Throughout, we shall set $c=G=\hbar=k=1$.

\section{Rigorous solution in deformed H\v{o}rava-Lifshitz gravity}
\vspace*{0.3cm}
\parindent=10pt
In the H\v{o}rava theory, a deformed action of the non-relativistic
renormalizable gravitational theory is given by \cite{KS}
\begin{eqnarray}
S_{HL}&=&\int dtd^3x \Big({\cal L}_0 + \tilde{{\cal L}}_1\Big)\nonumber\\
{\cal L}_0 &=& \sqrt{g}N\left\{\frac{2}{\kappa^2}(K_{ij}K^{ij}
\label{2.1}-\lambda K^2)+\frac{\kappa^2\mu^2(\Lambda_W R
  -3\Lambda_W^2)}{8(1-3\lambda)}\right\}\\ \tilde{{\cal L}}_1&=&
\sqrt{g}N\left\{\frac{\kappa^2\mu^2 (1-4\lambda)}{32(1-3\lambda)}R^2
-\frac{\kappa^2}{2w^4} \left(C_{ij} -\frac{\mu w^2}{2}R_{ij}\right)
\left(C^{ij} -\frac{\mu w^2}{2}R^{ij}\right) +\mu^4R \right\},
\label{2.2}
\end{eqnarray}
where $\kappa^2$, $\lambda$, $\mu$, $w$ and $\Lambda _W$ are
constant parameters, $K_{ij}$ is extrinsic curvature in the (3 +
1)-dimensional ADM formalism
\begin{eqnarray}
K_{ij}=\frac{1}{2N}\left(\dot{g}_{ij}-\nabla_i
N_j-\nabla_jN_i\right), \label{2.3}
\end{eqnarray}
and $C_{ij}$ is the Cotton tensor
\begin{eqnarray}
 C^{ij}=\epsilon^{ik\ell}\nabla_k
\left(R^{(3)j}{}_\ell-\frac{1}{4}R^{(3)}
 \delta^j_\ell\right). \label{2.4}
\end{eqnarray}
Taking the $\Lambda_W\rightarrow0$ limit and letting $\lambda=1$, it
was found that the speed of light, the Newton constant are described
by the following relations \cite{KS}
\begin{equation}
c^2= \frac{\kappa^2 \mu^4}{2}, \quad G = \frac{\kappa^2}{32 \pi c}
\label{2.5}
\end{equation}

A static and asymptotically flat KS black hole was found in
\cite{KS} which has the following form
\begin{eqnarray}
ds^2=-f(r)dt^2+\frac{dr^2}{f(r)}+r^2 (d\theta^2+\sin^2\theta d\phi^2
)\label{2.6}
\end{eqnarray}
with
\begin{eqnarray}
f(r)=1 + \omega r^2- \sqrt{r(\omega^2 r^3 + 4\omega M)}\label{2.7}
\end{eqnarray}
where $ \omega=\frac{16\mu^2}{\kappa^2} $ and M is an integration
constant related to the mass of the KS black hole.

The outer (inner) horizons are given by
\begin{eqnarray}
r_{\pm}=M\pm\sqrt{M^2-\frac{1}{2\omega}}\label{2.9}
\end{eqnarray}

The corresponding Hawking temperature and heat capacity are
\cite{MYU}
\begin{eqnarray}
T_H=\frac{1}{4\pi}\partial_r f(r)\mid_{r=r_+}=\frac{\omega
(r_+-M)}{2\pi(1+\omega r_+^2)}=\frac{2\omega r_+^2-1}{8\pi(\omega
r_+^3+r_+)}\label{2.10}
\end{eqnarray}
\begin{eqnarray}
C=\frac{\partial M}{\partial T}=-\frac{2\pi}{\omega}\frac{(1+\omega
r_+^2)^2(2\omega r_+^2-1)}{2\omega^2r_+^4-5\omega
r_+^2-1}\label{2.11}
\end{eqnarray}

\section{equilibrium fluctuation for KS black hole in deformed gravity}
\label{sec: intro2} \vspace*{0.3cm}
\parindent=10pt

In thermodynamical fluctuational theory, we always express
fluctuations of thermodynamical quantities in an equilibrium system
as some mean square fluctuations. If the mean square fluctuations
are divergent at a point, we call the point as a phase transition
point for the thermodynamical system. If the entropy of the system
is continuous at the point, we name the phase transition as the
SOPT. Therefore, we can look for the SOPT point in a thermodynamical
system according to singularity of fluctuations of the
thermodynamical quantities and  continuity of the entropy.

We suppose  that the KS black hole \cite{KS} is initially at
temperature $T_o$ (the point $O$ is arbitrary).  The variation of
temperature due to thermal fluctuation of the mass is $\Delta
\epsilon T_o$, where $ \mid \Delta \epsilon \mid \leq B$ and $B$
being a sufficiently small constant. From the first law of the
thermodynamics we have
\begin{eqnarray}
\Delta M=T_o(1+\Delta \epsilon) \Delta S \approx T_o \Delta
S\label{3.1}
\end{eqnarray}
where $T_o$ is determined by (\ref{2.10}).

The fluctuation probability $p$ is
\begin{eqnarray}
p\propto \exp{\Delta S}\sim \exp{(\frac{\Delta M}{T_o})}.
\label{3.2}
\end{eqnarray}
We know from the definition of heat  capacity that
\begin{eqnarray}
 \Delta M&=&CT_o\Delta \epsilon \label{3.3}.
\end{eqnarray}
where heat capacity $C$ is determined by (\ref{2.11}).

Because $ \mid \Delta \epsilon \mid \leq B$, the variation of the
mass $\Delta M$ may be expressed as\cite{Lau94b}
\begin{eqnarray}
 \mid \Delta M\mid &\leq &\mid C\mid T_oB \label{3.4}.
\end{eqnarray}
With Eqs. (\ref{3.2}), (\ref{3.4}) and the condition of normalized
\begin{eqnarray}
\int^{B\mid C\mid T_o} _{-B\mid C\mid T_o} \ \ p \ \ d(\Delta
M)=1,\label{3.5}
\end{eqnarray}
 we find that the
fluctuation probability $p$ can be rewritten as
\begin{eqnarray}
p=\frac{1}{2T_o\sinh{(B\mid C\mid)}}\exp{(\frac{\Delta
M}{T_o})}\label{3.6}.
\end{eqnarray}
Making use of Eqs.(\ref{3.4}) and (\ref{3.6}), we can get the  mean
square fluctuation of $M$
\begin{eqnarray}
\langle(\Delta M)^2\rangle &=& \frac{1}{2T_o\sinh{(B\mid
C\mid)}}\int^{B\mid C\mid T_o}
_{-B\mid C\mid T_o} (\Delta M)^2\exp{(\frac{\Delta M}{T_o})}d(\Delta M) \nonumber\\
&=&T^2_o\Big(2+B^2C^2-2B\mid C\mid\coth (B\mid
C\mid)\Big)\label{3.7}.
\end{eqnarray}
In the same way, we obtain
\begin{eqnarray}
\langle(\Delta S)^2\rangle &=&\frac{1}{T^2_o}
\langle(\Delta M)^2\rangle= 2+B^2C^2-2B\mid C\mid\coth (B\mid C\mid)\label{3.8},\\
\langle(\Delta T)^2\rangle &=&\frac{1}{\mid C\mid ^2}\langle(\Delta
M)^2\rangle =\frac{T^2_o}{C^2}\Big(2+B^2C^2-2B\mid C\mid\coth(B\mid
C\mid)\Big)\label{3.9}.
\end{eqnarray}
We observe from Eqs. (\ref{2.10}), (\ref{2.11}), (\ref{3.7}) and
(\ref{3.8})
  that $\langle(\Delta S)^2\rangle \rightarrow
\infty$ and $\langle(\Delta M)^2\rangle \rightarrow \infty$  when
$C\rightarrow\infty$. In other words, at the point $C\rightarrow
\infty $, even if we add the small perturbation of $ M $ to black
hole,  the fluctuations of mass and entropy of the KS black hole
will become so large that we can't describe it with a normal
equilibrium state. Thus, when the heat capacity changes from
positive to negative, we think that the KS black hole changes from a
phase into another phase and the critical point lies at
$\frac{\sqrt{5+\sqrt{33}}(\sqrt{33}-1)} {16\sqrt{\omega}} $. It is a
SOPT point because the entropy of the KS black hole is continuous at
this critical point. On the other hand, when $r_+\rightarrow r_-$,
Eqs. (\ref{3.7}) and (\ref{3.8}) lead to $\langle(\Delta S)^2\rangle
\rightarrow 0$, $\langle(\Delta M)^2\rangle \rightarrow 0$. It means
that when the KS black hole has a small deviation from equilibrium
state, all thermodynamic fluctuations goes to zero under the
condition that $r_+\rightarrow r_-$, and then the KS black hole will
come back to the initial equilibrium state. So we think the SOPT of
the KS black hole in the deformed H\v{o}rava-Lifshitz gravity
doesn't occur as $r_+\rightarrow r_-$ but $C\rightarrow\infty$.

\section{non-equilibrium fluctuation for KS black hole in deformed gravity}
\vspace*{0.3cm}
\parindent=10pt

In Landau and Lifshitz fluctuational theory
\cite{Landau-Lifshitz}, thermodynamical parameters and  entropy in
an arbitrary thermodynamical system will change as variation of
time. In order to describe the variation of the entropy, we
introduce a new physical quantity called the rate of entropy
production which can be expressed as
\begin{eqnarray}
\dot{S}=\sum_i X_i\dot{\chi_i} \label{4.1},
\end{eqnarray}
where $\dot{\chi_i}$ is flux of a given thermodynamic quantity
$\chi_i$, and $ X_i$ is thermodynamical force which is the conjugate
quantity of the thermodynamical flux $\chi_i$. In the KS black hole,
its mass parameter $M$, will be changed because of radiation of the
black hole in the real physic background. The rates $\dot{M}$ for
all emission processes can be written as
\begin{eqnarray}
\dot{M}&=&-fM^{-2}, \label{4.2}
\end{eqnarray}
where $f$ is dimensionless coefficient and can be calculated
numerically.

The rate of entropy production of the KS black hole \cite{KS} is
written as specifically
\begin{eqnarray}
\dot{S}&=&X_M\dot{M}\label{4.3}.
\end{eqnarray}with
\begin{eqnarray}
X_M&=&\frac{\partial S}{\partial M}=\frac{1}{T},\label{4.4}
\end{eqnarray}

In a general fluctuation-dissipative process, the flux
$\dot{\chi_i}$ is given by
\begin{eqnarray}
\dot{\chi_i}=\sum_k \Gamma_{ik}X_k \label{4.5},
\end{eqnarray}
where the quantity $\Gamma_{ik}$ is the phenomenological transport
coefficient which is defined as
\begin{eqnarray}
\Gamma_{ik}=\frac{\partial{\dot{\chi_i}}}{\partial{X_k}}\label{4.6}.
\end{eqnarray}
In non-equilibrium thermodynamic theory, we always make use of a
second moment to describe the fluctuations of a thermodynamic
quantity in a system. The second moments in the fluctuations of
the flux obey
\begin{eqnarray}
\langle \delta \dot{\chi_i} \delta\dot{\chi_k} \rangle
=(\Gamma_{ik}+\Gamma_{ki})\label{4.7},
\end{eqnarray}
where angular bracket denotes the mean value with respect to the
steady state, and the fluctuation $\delta \dot{\chi_i}$ is
spontaneous deviation from the steady-state value $\langle
\dot{\chi_i}\rangle$. If the second moment tends to infinite at a
certain point, we think a phase transition occurs in the system.

From Eqs. (\ref{4.2}), (\ref{4.4}) and (\ref{4.6}), we can get the
phenomenological transport coefficients
\begin{eqnarray} \Gamma_{MM}&=&-\frac{2fT^2}{M^3}(\frac{\partial
M}{\partial T})=-\frac{2fT^2C}{M^3},\label{4.8}.
\end{eqnarray}
Substituting (\ref{4.8}) into (\ref{4.7}), we find the second moment
for mass $M$
\begin{eqnarray}
\langle \delta \dot{M} \delta \dot{M}\rangle
=2\Gamma_{MM}=-\frac{4fT^2C}{M^3}\label{4.9} .
\end{eqnarray}
The second moments of entropy $S$ and temperature $T$ can be
 expressed as
\begin{eqnarray}
&\langle \delta\dot{S}\delta\dot{S}\rangle&=\frac{\langle \delta
\dot{M} \delta \dot{M}\rangle}{T^2}=-\frac{4fC}{M^3}\label{4.10},
\\ &\langle \delta \dot{T} \delta \dot{T}\rangle &=\frac{\langle
\delta
\dot{M} \delta \dot{M}\rangle}{C^2}=\frac{-4fT^2}{M^3C} \label{4.11},\\
&\langle \delta \dot{S} \delta \dot{T}\rangle &=\frac{\langle \delta
\dot{M} \delta \dot{M}\rangle}{T C}= \frac{-4fT}{M^3}\label{4.12}.
\end{eqnarray}

As $r_+\rightarrow r_-$, we find from Eqs. (\ref{4.9}),
(\ref{4.10}), (\ref{4.11}) and (\ref{4.12}) that
\begin{eqnarray}
\langle \delta \dot{M} \delta \dot{M}\rangle  \rightarrow 0, \ \ \ \
\langle \delta \dot{S} \delta \dot{S}\rangle \rightarrow 0, \ \ \ \
\ \langle \delta \dot{T} \delta \dot{T}\rangle \rightarrow 0, \ \ \
\ \ \langle \delta \dot{S} \delta \dot{T}\rangle \rightarrow
0.\label{4.13}
\end{eqnarray}
These equations show that all second moments of the KS black hole
tend to zero when $r_+\rightarrow r_-$, i.e., all thermodynamic
fluctuations vanish under this condition. Thus, the phase transition
does not occur in the KS black hole as $r_+\rightarrow r_-$.

On the other hand, when the heat capacity $C=\frac{\partial
M}{\partial T}\rightarrow \infty$, according to Eqs. (\ref{4.9}) and
(\ref{4.10}), we have
\begin{eqnarray}
&\langle \delta \dot{M} \delta \dot{M}\rangle &\rightarrow \infty,\nonumber\\
&\langle \delta \dot{S} \delta \dot{S}\rangle &\rightarrow
\infty.\label{4.14}
\end{eqnarray}
It is obvious that these second moments tend to infinite when heat
capacity is discontinuous. Thus, the thermal non-equilibrium
fluctuations  of some thermodynamical quantities are divergent.
According to the theory of thermal non-equilibrium fluctuation, we
can draw a conclusion that a phase transition of the KS black hole
occurs when $C\rightarrow \infty$. The phase transition is SOPT
since the entropy of the thermodynamical system is continuous.

\section{SUMMARY AND DISCUSSIONS}
\vspace*{0.3cm}
\parindent=10pt

In this paper, we investigate the SOPT of the KS black hole
\cite{KS} in the deformed H\v{o}rava-Lifshitz gravity by using the
thermal equilibrium and the thermal non-equilibrium fluctuation
methods, respectively. We find that as the heat capacity goes to
infinity the mean square fluctuations of mass and entropy in
equilibrium fluctuation diverges, so does the second moments of mass
and entropy in non-equilibrium fluctuation. However, under the
condition $r_+\rightarrow r_-$, all of fluctuations of thermodynamic
quantities in the both equilibrium and non-equilibrium fluctuations
vanish. Comparing to Schwarzschild black hole, although both of them
have one mass parameter, there are two phases ( one is
$\frac{1}{\sqrt{2\omega}}\leq M
<\frac{\sqrt{5+\sqrt{33}}(\sqrt{33}-1)} {16\sqrt{\omega}} $ and the
other one is $M> \frac{\sqrt{5+\sqrt{33}}(\sqrt{33}-1)}
{16\sqrt{\omega}} $) and the SOPT can taken place for the KS black
hole in the deformed H\v{o}rava-Lifshitz gravity. In this sense,
there is difference between the H\v{o}rava-Lifshitz gravity and the
usual general relativity.

\vspace*{1.5cm}
\begin{center}{ Acknowledgements}\end{center}
This work was supported by the National Natural Science
Foundation of China under Grant No 10875040;  a key project of the
National Natural Science Foundation of China under Grant No
10935013;  the National Basic Research of China under Grant No.
2010CB833004,  the Hunan Provincial Natural Science Foundation of
China under Grant No.  08JJ3010,   PCSIRT under Grant No.  IRT0964,
and the Construct Program  of the National Key Discipline. S. B. Chen's work was partially supported
by the National Natural Science Foundation of China under Grant
No.10875041; the Scientific Research Fund of Hunan Provincial
Education Department Grant No.07B043.

\end{document}